\documentclass{aa}

\usepackage[varg]{txfonts}
\usepackage{natbib,twoopt}
\usepackage[breaklinks=true]{hyperref} 
\bibpunct{(}{)}{;}{a}{}{,}             
\usepackage[dvipsnames]{xcolor}

\makeatletter
  \newcommandtwoopt{\citeads}[3][][]{\href{http://adsabs.harvard.edu/abs/#3}%
    {\def\hyper@linkstart##1##2{}%
     \let\hyper@linkend\@empty\citealp[#1][#2]{#3}}}
  \newcommandtwoopt{\citepads}[3][][]{\href{http://adsabs.harvard.edu/abs/#3}%
    {\def\hyper@linkstart##1##2{}%
     \let\hyper@linkend\@empty\citep[#1][#2]{#3}}}
  \newcommandtwoopt{\citetads}[3][][]{\href{http://adsabs.harvard.edu/abs/#3}%
    {\def\hyper@linkstart##1##2{}%
     \let\hyper@linkend\@empty\citet[#1][#2]{#3}}}
  \newcommandtwoopt{\citeyearads}[3][][]%
    {\href{http://adsabs.harvard.edu/abs/#3}
    {\def\hyper@linkstart##1##2{}%
     \let\hyper@linkend\@empty\citeyear[#1][#2]{#3}}}
    \renewcommand*\aa@pageof{, page \thepage{} of \pageref*{LastPage}}
\makeatother

\newcommand{\Inu}{I_{\nu}}
\newcommand{\Inue}{I^{\mathrm{e}}_{\nu}}
\newcommand{\Jnu}{J_{\nu}}
\newcommand{\Bnu}{B_{\nu}}
\newcommand{\Rnu}{R_{\nu}}
\newcommand{\Rnue}{R_{\nu}^{\mathrm{e}}}
\newcommand{\Dnu}{D_{\nu}}
\newcommand{\Dnue}{\Dnu(\Rnue)}
\newcommand{\psinu}{\psi_\nu}
\newcommand{\hnu}{\Vec{h}_\nu}
\newcommand{\Rin}{R_{\mathrm{in}}}
\newcommand{\Rout}{R_{\mathrm{out}}}
\newcommand{\Rstar}{R_{\star}}
\newcommand{\Tstar}{T_{\star}}
\newcommand{\Teff}{T_{\mathrm{eff}}}
\newcommand{\Tin}{T_{\mathrm{in}}}
\newcommand{\rs}{\vec{r}_\mathrm{S}}
\newcommand{\Hnu}{\Vec{H}_{\nu}}
\newcommand{\Fnuin}{F^{\mathrm{in}}_{\nu}}

\newcommand{\Knusca}{\kappa^{\mathrm{sca}}_{\nu}}
\newcommand{\Knuabs}{\kappa^{\mathrm{abs}}_{\nu}}
\newcommand{\Knuext}{\kappa^{\mathrm{ext}}_{\nu}}
\newcommand{\Cnusca}{C^{\mathrm{sca}}_{\nu}}
\newcommand{\Cnuabs}{C^{\mathrm{abs}}_{\nu}}
\newcommand{\Cnuext}{C^{\mathrm{ext}}_{\nu}}
\newcommand{\nhat}{\hat{n}}
\newcommand{\rhat}{\hat{r}}
\newcommand{\shat}{\hat{s}}
\DeclareMathOperator{\mean}{mean}

\begin{document}

\title{New boundary conditions for the approximate flux-limited diffusion radiative transfer in circumstellar environments}
\subtitle{Test case study for spherically symmetric envelopes}

\author{J. Perdigon \and G. Niccolini \and M. Faurobert}

\institute{Universit\'{e} C\^{o}te d’Azur (UCA), Observatoire de la C\^{o}te d’Azur (OCA), CNRS, Laboratoire Lagrange, Lagrange, France \\ \email{jeremy.perdigon@oca.eu}}

\date{Received <date>/Accepted <date>}

\abstract
{In order to constrain the models describing circumstellar environments, it is necessary to solve the radiative transfer equation in the presence of absorption and scattering, coupled with the equation for radiative equilibrium. However, solving this problem requires much CPU time, which makes the use of automatic minimisation procedures to characterise these environments challenging.}
{In this context, the use of approximate methods is of primary interest. One promising candidate method is the flux-limited diffusion (FLD), which recasts the radiative transfer problem into a non-linear diffusion equation. One important aspect for the accuracy of the method lies in the implementation of appropriate boundary conditions (BCs). We present new BCs for the FLD approximation in circumstellar environments that we apply here to spherically symmetric envelopes.}
{At the inner boundary, the entering flux (coming from the star and from the envelope itself) may be written in the FLD formalism and provides us with an adequate BC. At the free outer boundary, we used the FLD formalism to constrain the ratio of the mean radiation intensity over the emerging flux. In both cases we derived non-linear mixed BCs relating the surface values of the mean specific intensity and its gradient. We implemented these conditions and compared the results with previous benchmarks and the results of a Monte Carlo radiative transfer code. A comparison with results derived from BCs that were previously proposed in other contexts is presented as well.}
{For all the tested cases, the average relative difference with the benchmark results is below 2\% for the temperature profile and below 6\% for the corresponding spectral energy distribution or the emerging flux. We point out that the FLD method together with the new outer BC also allows us to derive an approximation for the emerging flux. This feature avoids additional formal solutions for the radiative transfer equation in a set of rays (ray-tracing computations).}
{The FLD approximation together with the proposed new BCs performs well and captures the main physical properties of the radiative equilibrium in spherical circumstellar envelopes.}


\titlerunning{New boundary conditions for the approximate FLD radiative transfer in circumstellar environments}
\authorrunning{J. Perdigon et al.}

\maketitle

\section{Introduction}

The study of circumstellar environments at different stages of stellar evolution is of crucial importance. These environments reflect the physical processes in action, from the star formation with the presence of accretion discs to late stages in the evolution, in which strong stellar winds shape the circumstellar envelopes. Observations at high angular resolution allow to probe and characterise the circumstellar material by determining densities, temperatures, abundances, velocity fields, etc. The exploitation of instruments such as the Multi AperTure mid-Infrared SpectroScopic Experiment \footnote{\href{http://www.eso.org/public/teles-instr/paranal-observatory/vlt/vlt-instr/matisse}{http://www.eso.org/public/teles-instr/paranal-observatory/vlt/vlt-instr/matisse}} (MATISSE), operating in the mid-IR, or the Atacama Large Millimeter Array \footnote{\href{http://www.almaobservatory.org}{http://www.almaobservatory.org}} (ALMA) in the sub-millimetric domain offer complementary views of these environments, that give access to regions close to the central star up to the outer regions using a multi-wavelength approach. 

Circumstellar matter is generally composed of a mixture of gas and dust particles that absorbs and scatters the incident stellar radiation. The envelope is then heated by the radiation, and a radiative equilibrium can be reached in which the envelope also emits radiation in the infrared domain.

In order to constrain the models describing circumstellar environments, it is necessary to solve the radiative transfer equation under the assumption of radiative equilibrium. Several numerical techniques exist to solve this problem in one, two, and three dimensions. The Monte Carlo method is popular because it can be adapted to any geometry and can handle many physical processes (\citeads[see]{2013ARA&A..51...63S} for a thorough review). However, solving the radiative transfer problem requires much CPU time, which makes the use of any automatic minimisation procedure to characterise these environments challenging. 

In this context, the use of approximate methods is of primary interest. One promising candidate is the flux-limited diffusion (FLD), introduced by \citetads{1981ApJ...248..321L} (L\&P hereafter). This description numerically simplifies the problem by recasting the radiative transfer equation into a non-linear diffusion equation for the mean specific intensity of the radiation field (see Sect.~\ref{sect:FLD}). 

Physically, the boundary conditions (BCs) for the radiative transfer equation are obtained from the known specific intensity incident on the surface of the object. However, the outgoing intensity is not known a priori and has to be obtained from the solution of the radiative transfer problem. It is thus not obvious to find a BC for the mean specific intensity at the surface of the object. A consequent theoretical work has already been done in finding satisfying BCs for the FLD method \citepads{1986JQSRT..36..325P, 1988JCoPh..75...73P}. These BCs were derived with the assumption that a boundary layer could be defined, which might not be true in astrophysical applications where the media may not be seen as an infinite half-space by the radiation for some frequencies. Furthermore, as far as we know, they have never been numerically tested in an astrophysical context. 

The FLD approximation has already been implemented in several astrophysical applications. \citetads{1995A&A...299..545S, 2002ApJ...569..846Y} used the FLD to solve the frequency-dependent radiative transfer to model protostellar discs and massive star formation, respectively. In these studies, the central star was treated as an additional source and the specific mean intensity $\Jnu$ at the outer edge of the media was set to be equal to the Planck function $\Bnu(T_\mathrm{out})$, with a prescribed temperature $T_\mathrm{out}$ at the boundary. Some improvements were made later, in the context of the radiation hydrodynamics problem for massive star formation (\citeads[see]{2010A&A...511A..81K, 2020A&A...635A..42M}). These more sophisticated hybrid codes split the radiation field into two components, the stellar and the dust component, where the FLD method only solves for the latest part. In this treatment, the Dirichlet boundary condition at the outer edge only applies to the dust component. This relies on the assumption that the dust temperature is known at the interface with the interstellar medium. In the problem we consider, the temperature at the outer boundary is not known a priori and must be derived as part of the solution to the radiative transfer problem coupled with the radiative equilibrium equation. In a non-grey problem, we need BCs that can properly handle several regimes of optical thicknesses, for different frequencies.

We present new BCs for the FLD theory in circumstellar environments that we tested and implemented in the case of spherically symmetric envelopes. The condition is derived from the prescription of (i) the incident flux, derived from an extended stellar source and the self-heating of the envelope at the inner boundary and from (ii) the ratio of the mean specific intensity over the radiative flux at the free outer boundary. We show that they both may be written as mixed BCs relating the mean specific intensity and its gradient at the surfaces of the envelope. They consequently lead to a more realistic description of the radiation field (compared to simple Dirichlet or von Neumann boundary conditions) while still remaining sufficiently easy to implement. As a by-product of our investigations, we also derived an approximate expression for the emergent flux at the free outer surface.

The paper is organised as follows: in Sect.~\ref{sect:FLD} we recall the bases of the FLD theory. In Sect.~\ref{sect:Boundary_Conditions} we present the new boundary conditions and in Sect.~\ref{sect:numerical} their numerical implementations. In Sect.~\ref{sect:1D_tests} we test the accuracy of our results by comparing the temperature profile in the envelope and the spectral energy distribution (SED) of the outgoing flux with the results of two radiative transfer codes, namely (i) DUSTY \citepads{1997MNRAS.287..799I}, which numerically solves the integral equation for the energy density, and (ii) a Monte Carlo (MC) radiative transfer code \citepads{2006A&A...456....1N}. Additionally, we compare the derived boundary conditions with the original boundary conditions of \citetads{1981ApJ...248..321L}. Finally, in Sect.~\ref{sect:conclusion}, we conclude and present some perspective for our future work.
\section{The Flux Limited Diffusion Theory} \label{sect:FLD}

In the following, we present the original work of L\&P and introduce the relevant background for the derivation of the BCs in Sect.~\ref{sect:Boundary_Conditions}. The position is denoted by $\vec{r}$, the direction of propagation by $\nhat$, and the frequency by the subscript $\nu$. The transport equation for the specific intensity $\Inu\left(\vec{r},\nhat,t\right)$ at the position $\vec{r}$ in the $\nhat$ direction with isotropic and coherent scattering is
\begin{equation}
    \frac{1}{c} \partial_t \, \Inu + \nhat.\vec{\nabla} \Inu = - \Knuext \Inu + \Knuabs \Bnu + \Knusca \Jnu \ . \label{eq:transport}
\end{equation}
$\Jnu = \Jnu \left(\vec{r},t\right)$ is the mean specific intensity, $\Bnu = \Bnu \left( T \left(\vec{r},t\right) \right)$ is the Planck function and $\Knuext$, $\Knuabs$ and $\Knusca$ are the extinction, absorption and scattering coefficients, respectively. The zeroth and first moments of the specific intensity, namely $\Jnu$ and $\Hnu$, are defined as
\begin{equation}
    \Jnu = \frac{1}{4\pi}\int\limits_{4\pi}^{} \Inu \ d^2\nhat \ , \ \Hnu = \frac{1}{4\pi}\int\limits_{4\pi}^{} \Inu \ \hat{n} \ d^2\nhat \ , \label{eq:moments}
\end{equation}
where the integration is performed over all directions. These quantities are linked by the zeroth moment of Eq.~(\ref{eq:transport}),
\begin{equation}
    \frac{1}{c} \partial_t \ \Jnu + \vec{\nabla}.\Hnu = \Knuabs\left( \Bnu - \Jnu \right) \ . \label{eq:moment_transport}
\end{equation}
The FLD approximation is a closure of the system of the moment equations by expressing $\Hnu$ as a function of $\Jnu$. This is done by expressing the specific intensity $\Inu$ as a function of the mean specific intensity $\Jnu$,
\begin{equation}
    \Inu = \Jnu \ \psinu \left(\vec{r},\hat{n},t\right) \ , \ \Hnu = \Jnu \ \hnu \left(\vec{r},t\right) \ , \  \frac{1}{4\pi}\int\limits_{4\pi}^{} \psinu \ d^2\nhat = 1 \ , \label{eq:I_H_FLD}
\end{equation}
where $\hnu$ is the normalised flux and is expressed as
\begin{equation}
    \hnu \left(\vec{r},t\right) = \frac{1}{4\pi}\int\limits_{4\pi}^{} \nhat \, \psinu\left(\vec{r},\nhat,t\right)  \ d^2\nhat \ . \label{eq:hnu}
\end{equation}
The $\psinu$ function is called the normalised intensity and quantifies the anisotropy of the radiation field $\Inu$. In the optically thin and thick limits, this function reduces to a Dirac distribution and a constant, respectively. The FLD approximation consists of assuming that the anisotropy of the radiation field is a conserved quantity, yielding the expression for $\psinu$,
\begin{equation}
\psinu\left(\vec{r},\nhat,t\right) =  \frac{1}{1+  \left( \hnu- \nhat\right) .\vec{R}_\nu} \ , \label{eq:psinu} 
\end{equation}
with 
\begin{equation}
\vec{R}_\nu\left(\vec{r},t\right) = \frac{- \vec{\nabla}\Jnu}{\omega_\nu~\Knuext ~\Jnu} ~~~~ , ~~~~ \omega_\nu = \frac{\Knuabs \Bnu + \Knusca \Jnu}{\Knuext \Jnu} \ .  \label{eq:R}
\end{equation}
The quantity denoted by $\Rnu$ plays a key role in the description of the local radiation field in the medium. It expresses the ratio of the (effective) mean free path over the characteristic length of the variation of the mean specific intensity. Consequently, the limits $R_\nu \gg 1$ and $R_\nu \ll 1$ correspond to the optical thin and thick regimes, respectively. The quantity $\omega_\nu$ is called the effective albedo. It is equal to unity in the absence of true absorption ($\Knuabs=0$). L\&P showed that, using Eq.~(\ref{eq:psinu}) in Eq.~(\ref{eq:hnu}), $\hnu$ is proportional to  $\vec{R}_\nu$,
\begin{equation}
\hnu = \lambda\left(\Rnu\right) \vec{R}_\nu ~~~~ , ~~~~ \lambda\left(\Rnu\right) = \frac{1}{\Rnu}\left( \frac{1}{\tanh{\Rnu}} - \frac{1}{\Rnu} \right) \ , \label{eq:h}
\end{equation}
where $\Rnu$ is the norm of $\vec{R}_\nu$ and $\lambda\left(\Rnu\right)$ is the 'flux-limiting' parameter. Finally, using Eq.~(\ref{eq:h}) in Eq.~(\ref{eq:I_H_FLD}) allows  Eq.~(\ref{eq:moment_transport}) to be rewritten as a non-linear diffusion equation for the mean specific intensity
\begin{equation}
    \frac{1}{c}\partial_t \ \Jnu - \vec{\nabla}.\left( \Dnu \vec{\nabla} \Jnu \right) = \Knuabs \left( \Bnu - \Jnu \right) \ , \label{eq:FLD}
\end{equation}
with the non-linear diffusion coefficient 
\begin{equation}
 \Dnu = \frac{\lambda\left(\Rnu\right)}{\omega_\nu \Knuext} \ .  \label{eq:Dnu} 
\end{equation}
We note that in the FLD approach, the radiative net flux $\Hnu$ is related to the gradient of the mean specific intensity $\Jnu$ by $\Hnu = -\Dnu \vec{\nabla}\Jnu$. It shows some similarities with the Fick law, which applies in the stellar interior, but here the diffusion coefficient depends on a non-linear way on the mean specific intensity. In the optically thick regime, the FLD approximation reduces to a linear diffusion equation, whereas in the optically thin regime, it reduces to an advection equation, as expected.
\section{Boundary conditions} \label{sect:Boundary_Conditions}

In this section, we specify the time-independent BCs that were implemented for the FLD Eq.~(\ref{eq:FLD}). They are defined at the specific boundaries of a circumstellar shell, namely an inner spherical cavity that is illuminated by an enclosed star, and an outer boundary with no incoming radiation. 

The BCs for the radiative transfer equation Eq.~(\ref{eq:transport}) would be obtained by specifying the value of the incident specific intensity $\Inu$ on the considered surface. In the framework of the FLD approximation, we face two problems: (i) The FLD equation applies to the mean specific intensity, whereas the physical constraint is on the ingoing specific intensity at the boundaries, and (ii) we also explained in Sect.~\ref{sect:FLD} that the FLD approach implies a specific angular dependence of $\Inu$, given by the function $\psinu$ (see Eq.~(\ref{eq:psinu})). This specific dependence would not be consistent with any arbitrary value of the incident radiation field at the surface. Consequently, the actual BCs for the radiative transfer equation are in general incompatible with the FLD solution.

This is expected because the FLD approximation is in principle not valid close to the boundaries of objects. \citetads{1988JCoPh..75...73P} derived the BCs for the FLD equation from the decomposition of the radiative transfer problem into an interior problem described by the FLD equation and an additional boundary layer term. The match of the interior and boundary layer solutions yields the BCs for the FLD equation. A relation of the surface values of the mean specific intensity and its gradient is obtained, but with rather involved coefficients depending on integrals of the Chandrasekhar $H$ function and on the surface value of $\Rnu$. In the literature, Dirichlet BCs are often used with a prescribed value for the temperature at the outer boundary of the medium. As already pointed out, the surface temperature is not known a priori and must be derived from the solution of the radiative transfer problem. 

Here, we propose to impose BCs for the zeroth and the first angular moments of the radiation field, in a form that is compatible with the FLD approach, in order to ensure a smooth match with the interior FLD solution. We obtained mixed Robin-type BCs that relate the surface values of the mean specific intensity and its gradient, but the coefficients are quite simple analytical functions of the surface value of $\Rnu$. 

\subsection{Inner spherical cavity}  \label{subsect:inner_cavity}

\subsubsection{Expression of the incident flux in the FLD formalism}

One example of physical significance is to write a condition on the flux $\Fnuin$ entering a boundary surface at the position $\Vec{r}_S$,
\begin{equation}
    \Fnuin(\rs)  = \Jnu(\rs) \int\limits_{\shat.\nhat\leq 0}^{} \shat.\nhat \ \psinu(\rs,\nhat) \ d^2\nhat \ . \label{eq:boundary_equation_1}
\end{equation}
Here, $\shat$ is the outward normal vector to the surface of the envelope ($\shat = - \hat{r}$ at the inner edge). The right-hand side (RHS) of Eq.~(\ref{eq:boundary_equation_1}) is the expression of an incoming flux in the FLD formalism. Eq.~(\ref{eq:boundary_equation_1}) is the general form of the inner BC without any assumption on the geometry of the problem. The integral can be performed analytically using Eq.~(\ref{eq:psinu}) for $\psinu$ if we assume that the vector $\vec{R}_\nu$ is normal to the surface ($\vec{R}_\nu = \pm \, R_\nu \ \rhat$), which is exact for spherically symmetric problems. The equation can then be rewritten as a non-linear mixed BC,
\begin{equation}
    \Fnuin(\rs)  = \pi \left( \alpha_\nu \ \Jnu(\rs) + \beta_\nu \ \shat.\vec{\nabla} \Jnu|_{\vec{r}=\rs} \right) \label{eq:BCg}
\end{equation}
with
\begin{equation}
\alpha_\nu  =  \left. 2 \frac{\ln{(\cosh{\Rnu}})}{\Rnu\tanh{\Rnu}}\right|_{\vec{r}=\rs} ~~~~,~~~~ \beta_\nu   =   2 \Dnu|_{\vec{r}=\rs}. \label{eq:alpha}
\end{equation}
This BC may be regarded as an implicit relation of the surface values of $\Jnu$ and $\Rnu$ that yields an FLD solution compatible with the given incident flux. It is interesting to note that although Eq.~(\ref{eq:BCg}) is different in its coefficients and construction from the original BC Eqs.~($56$) and ($66$) in L\&P, they are analytically equivalent. In L\&P, the coefficients were derived to give an exact transport result for the case of a source-free ($\Bnu = 0$) half-space media, with a constant $\Knuext$, $\omega_\nu$ and a particular incident intensity distribution $\Gamma(\rs,\mu)$ of the form
\begin{equation}
    \Gamma(\rs,\mu) = \frac{1}{\coth(\Rnu(\rs)) - \mu} = \Rnu(\rs) \  \psinu(\rs,\mu) \ . \label{eq:Gamma}
\end{equation}
This specific form is proportional to the angular dependence in the FLD formalism. The correspondence between $\alpha_\nu$ in Eq.~(\ref{eq:alpha}) and Eq.~($56$) in L\&P is
\begin{equation}
    \gamma = \frac{\alpha_\nu - 2 \lambda(\Rnu) \shat.\vec{R}_\nu}{2 - 4 \lambda(\Rnu) \shat.\vec{R}_\nu} \ .
\end{equation}
Hence, specifying the FLD flux at the boundaries of a spherically symmetric domain will actually give the same BC as is obtained by solving the exact transport result of a source-free ($\Bnu = 0$) half-space media, with a constant $\Knuext$, $\omega_\nu$ and a particular incident intensity distribution $\Gamma(\rs,\mu)$, given by Eq.~(\ref{eq:Gamma}). We note that this equivalence only applies for the spherical and planar symmetric systems and no longer holds when we specify Eq.~(\ref{eq:boundary_equation_1}) for other configurations, where the radiative flux is not orthogonal to the boundary surface.

\subsubsection{Incident flux from an extended source and the envelope} 

\begin{figure}
    \centering
    \resizebox{\hsize}{!}{\includegraphics{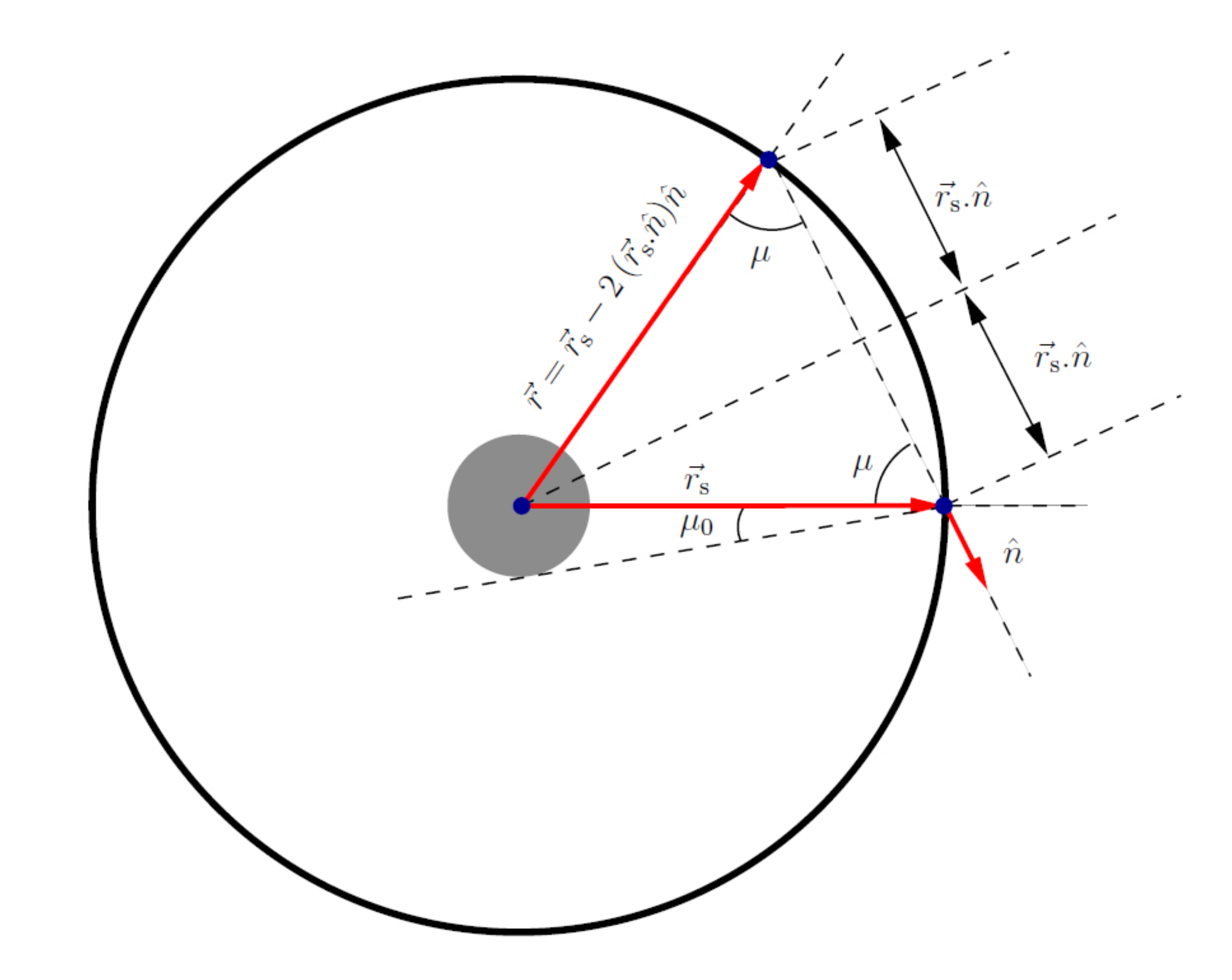}}
    \caption{Geometry of an inner spherical cavity (solid black line) illuminated by a central star (grey disc).}
    \label{fig:inner_cavity}
\end{figure}

We want to write the BC at an inner spherical cavity located at a distance $\Rin$ from the centre of a star of radius $\Rstar$ ($\vec{r}_S = \Rin \ \hat{r}$) and surface temperature $\Tstar$. For this, we need to specify the flux $\Fnuin$. We have two contributions,
\begin{equation}
    \Fnuin = \int\limits_{\shat.\nhat \leq 0}^{} \shat.\nhat \left[ \Bnu(\Tstar) + \Inue \left(\vec{r},-\nhat\right) \right] ~ d^2\nhat \ .
\end{equation}
The first contribution, denoted by $\Bnu(\Tstar)$ comes from the star and the other from the inner boundary itself and is expressed as $\Inue \left(\vec{r},-\nhat\right)=J_\nu \left( \vec{r}\right) \psinu\left(\vec{r},-\nhat\right)$. As shown by Fig.\ref{fig:inner_cavity}, the vector $\vec{r} = \rs - 2 ( \rs . \nhat) \ \nhat$ corresponds to the opposite point at the inner boundary, along $\nhat$. Because of this dependence, this BC is no longer local by nature and cannot be expressed in a closed form, except in spherical symmetry where $\Jnu(\vec{r}) = \Jnu(\rs)$ and $\psinu(\vec{r},- \nhat) = \psinu(\rs,-\nhat)$. To perform the angular integration, we aligned the $n_z$ axis with the unitary vector $\hat{r}$. For the star, the integration on $\mu = \cos(\theta)$ ($\theta$ being the angle between $n_z$ and $\nhat$) spans from $\mu_0=\sqrt{1-\left(\Rstar/\Rin\right)^2}$ to $1$, and for the inner cavity, it spans from $0$ to $\mu_0$,
\begin{equation}
    \Fnuin = 2 \pi \left( \int\limits_{\mu_0}^{1} \mu \ B_\nu(\Tstar) \ d\mu + J_\nu(\vec{r}_S) \int\limits_{0}^{\mu_0} \mu \ \psinu\left(\vec{r}_S,-\mu\right) \ d\mu \right) \ .
\end{equation}
The incident flux $\Fnuin$ is then expressed as,
\begin{equation}
    \Fnuin =  \pi \left[ \left(\frac{\Rstar}{\Rin}\right)^2 B_\nu(\Tstar) + \gamma_\nu ~ J_\nu\left( \vec{r}_S \right) \right] \ ,
\end{equation}
with,
\begin{equation}
\gamma_\nu = \left.\frac{2 \left[\mu_0\tanh{R_\nu}-\ln{\left(1+\mu_0\tanh{R_\nu}\right)} \right]}{R_\nu\tanh{R_\nu}}\right|_{\vec{r} = \vec{r}_S} \ . \label{eq:gamma}    
\end{equation}
The inner BC for the FLD equation in the case of an inner spherical cavity enclosing a star is
\begin{equation}
 \left( \alpha_\nu - \gamma_\nu  \right) ~ J_\nu\left( \Rin\right) + \beta_\nu ~ \hat{s}.\vec{\nabla} J_\nu|_{r = R_{in}} = \left(\frac{\Rstar}{\Rin}\right)^2 B_\nu(\Tstar) \ , \label{eq:BC_in}
\end{equation}
with $\alpha_\nu$ and $\beta_\nu$, the quantities defined by Eq.~(\ref{eq:alpha}). 

We can analytically write the FLD solution with this inner boundary condition in the limit where the envelope is optically thin ($\Rnu \gg 1$) and compare it with the known analytical solution for the dilution of the mean specific intensity in free space,
\begin{equation}
    \Jnu = \frac{\Bnu(\Tstar)}{2} \left(  1-\sqrt{1-\left(\frac{\Rstar}{r}\right)^2} \right) \ . \label{eq:v_field}
\end{equation}
In the optically thin limit, the FLD Eq.~(\ref{eq:FLD}) and the BC Eq.~(\ref{eq:BC_in}) reduce to  
\begin{equation}
    r^2 \Jnu = \text{const} = \Rin^2 \Jnu\left( \Rin \right) \ , \ \Jnu\left( \Rin \right) = \frac{1}{4} \left(\frac{\Rstar}{\Rin}\right)^2 B_\nu(\Tstar) \ ,
\end{equation}
respectively. The solution of the FLD equation, in this limit  is then
\begin{equation}
    \Jnu = \frac{1}{4} \left(\frac{\Rstar}{r}\right)^2 \Bnu(\Tstar) \ ,\label{eq:FLD_thin} 
\end{equation} 
in agreement with Eq.~(\ref{eq:v_field}) when $\left( \Rstar / r \right)^2 \ll 1$. The relative difference in temperature between Eq.~(\ref{eq:v_field}) and the FLD solution Eq.~(\ref{eq:FLD_thin}) is $\approx 15 \ \%$ at the star surface, $1 \ \%$ at $r \approx 2.5 \ \Rstar$. For $r \ga 10 \ \Rstar$ (values for our test cases presented in Sect.~\ref{sect:1D_tests}), this difference becomes negligible ($ \la 0.06 \ \%$).

\subsection{Outer boundary} \label{subsect:outerBC}

For the inner cavity, we would like to impose that no incoming radiation enters the external shell of the envelope ($F_\nu^\mathrm{in} = 0$). However, in the FLD formalism, the angular dependence of the radiation is given by the specific form of the $\psinu$ function (Eq.~(\ref{eq:psinu})). The incoming radiation vanishes only when $\Rnu$ becomes infinite, which also results in a sharp-peaked distribution for the emerging specific intensity. This is not physically realistic, and this inconsistency is expected because the FLD method is rigorously not valid close to the surface of the object. Another approach is then required to describe the behaviour of the radiation on the external edge. Inspired by L\&P and \citetads{1986JQSRT..36..325P}, we seek a BC in the form of a closure relation between the mean specific intensity and the radiation flux at the surface, that is,
\begin{equation}
    J_\nu(\Rout) - \zeta_\nu \ \hat{s}.\vec{H_\nu}(\Rout) = 0 \ ,
\end{equation}
with $\zeta_\nu$, a coefficient we need to determine. At the outer edge and without incoming radiation, $\zeta_\nu$ has to be understood as the ratio of the energy density over the emerging flux. This ratio can be expressed in spherical symmetry as
\begin{equation}
    \zeta_\nu = \frac{\int_{0}^{1} \Inu ~ d\mu}{\int_{0}^{1}\mu ~\Inu ~ d\mu}\ .
\end{equation}
It depends on the anisotropy of the emergent radiation field. In the optically thin limit, the radiation field is along the $\shat$ direction (spherical symmetry), thus $I_\nu \propto \delta(\mu-1)$ and hence $\zeta_\nu \xrightarrow[]{} 1$. In the diffusion regime where the emergent field is isotropic, $\zeta_\nu \xrightarrow[]{} 2$. In the original study of L\&P, $\zeta_\nu$ was chosen to be equal to $2$, which means that it correctly describes the optically thick cases where $\Rnu \ll 1$. If the BC is to correctly describe different optical regimes, we need $\zeta_\nu$ to be a function of $\Rnu$. In the framework of the FLD approximation, $\Inu = \Jnu \ \psinu$ and the specific angular dependence of $\Inu$, given by $\psinu$, is used to compute the surface value of $\zeta_\nu$,
\begin{equation}
    \zeta_\nu = \zeta(\Rnu) = \frac{2 + \alpha(R_\nu) \tanh(R_\nu)}{\alpha(R_\nu) + 2 \lambda(R_\nu)R_\nu} \ . \label{eq:zeta}
\end{equation}
In the two limits ($\Rnu \gg 1$ and $\Rnu \ll 1$), we recover $\zeta_\nu \xrightarrow[]{} 1$ and $\zeta_\nu \xrightarrow[]{} 2$. We note that the latter limit reduces to the boundary condition Eq.~($56$) in L\&P. Now we need to specify the value of $\Rnu$ at the external boundary of the envelope. The behaviour of the radiation at this interface is only dictated by the interior solution because there is no incoming radiation. Consequently, to ensure a smooth match of the BC and the interior FLD solution, we used a second-order extrapolated value for $\Rnu(\Rout) = \Rnue$ (see Eq.~\ref{eq:Rnuextrapolation}). Because we prescribe a value for $\Rnu$ at the external edge, $\Rnue$ is also used to compute the non-linear diffusion coefficient $\Dnu(\Rout)= \Dnue = \lambda(\Rnue) / \omega_\nu \Knuext $ and the coefficients of the numerical scheme Eq.(\ref{eq:coeff}) on the external edge. The outer BC without incoming radiation that we implemented is then
\begin{equation}
    \Jnu(\Rout) + \zeta(\Rnue) \ \Dnue \ \hat{s}.\left.\vec{\nabla}\Jnu \right|_{r=\Rout} = 0 \ , \label{eq:BC_out}
\end{equation}
where we have expressed $\Hnu(\Rout) = - \Dnue \ \vec{\nabla}\left.\Jnu\right|_{r=\Rout}$. In Sect.~\ref{subsect:FLD_vs_TAPAS} we compare this BC in an astrophysical application with respect to the original BCs of L\&P.

\subsection{Approximation for the emergent flux}

We used an extrapolation of the non-linear diffusion coefficient $\Dnu(\Rnue)$ to relate the flux at the external edge to the gradient of the mean specific intensity. As there is no incident flux, this provides us with an approximate expression for the emergent flux, given by
\begin{equation}
    \vec{F}_\nu^\mathrm{out} = 4\pi \Hnu(\Rout) = - 4\pi \  \Dnue \ \left. \vec{\nabla} \Jnu \right|_{r=\Rout} \ .
    \label{eq:approxflux}
\end{equation}
This approximation is tested in Sect.~\ref{sect:1D_tests}. It reproduces the results of different radiative transfer codes that solve the full transfer equation very well, as shown in Fig.~\ref{fig:FLD_vs_Ivezic} and Fig.~\ref{fig:FLD_vs_TAPAS}.

\subsection{Radiative equilibrium and warming of the stellar surface}\label{subsect:warming}

Because of the geometric extension of the star (see Fig.~\ref{fig:inner_cavity}), there is part of the radiation that emerges from the envelope that falls back onto the star,
\begin{equation}
\begin{split}
     F_\nu^{\mathrm{fall}}(\Rin) = 2 \pi \Jnu (\Rin) \int\limits_{-1}^{-\mu_0} \mu \ \psi_\nu \left(\Rin,\mu\right) \ d\mu \\
     = \frac{2 \pi \Jnu(\Rin)}{\Rnu \tanh{\Rnu}} \left[ \ln{\left(\frac{1+\tanh{\Rnu}}{1+\mu_0\tanh{\Rnu}}\right)} - \left(1-\mu_0\right)\tanh{\Rnu} \right] \ , \label{eq:Ffall}
\end{split}
\end{equation}
with $\Rnu$ being evaluated at $r = \Rin$. We used the same conventions as in Sect.~\ref{subsect:inner_cavity} to perform the angle integration. This part of the flux is hidden from the rest of the envelope, and the radiative equilibrium inside the cavity leads to a warming of the stellar surface \citepads{2003A&A...399..703N}. This effect can be quite dramatic, and reach up to $30 \ \%$ of the total stellar luminosity that is obscured for optically-thick grey shell, as in the test case presented in Sect.~\ref{subsect:FLD_vs_TAPAS}. This has to be taken into account to properly ensure the radiative equilibrium condition throughout the full space from the stellar surface to the outer boundary of the envelope. To fulfil the radiative equilibrium condition at the stellar surface we write
\begin{equation}
    \sigma \Teff^4 = \sigma \Tstar^4 + \left(\frac{\Rin}{\Rstar} \right)^2 \int\limits_{0}^{\infty}  F_\nu^{\mathrm{fall}}(\Rin) \ d\nu  \ .  \label{eq:teff_update}
\end{equation}
We imposed a fixed value for $\Teff$, and the temperature of the star $\Tstar$ was updated accordingly.
\section{Numerical implementation} \label{sect:numerical}

\noindent The FLD Eq.~(\ref{eq:FLD}) is a non-linear diffusion equation that has to be solved numerically for each point of space, time, and frequency. In the following, we limit ourselves to the 1D time-independent FLD equation,
\begin{equation}
    \frac{1}{r^2}\partial_r \left( r^2 D_\nu \ \partial_r J_\nu \right) = \Knuabs \left(J_\nu - B_\nu \right) \ .  \label{eq:1D_FLD}
\end{equation}
Here, $r$ denotes the radial variable from the centre of the envelope. An additional constraint to Eq.~(\ref{eq:1D_FLD}) is given by the equation of the radiative equilibrium
\begin{equation}
    \int\limits_{0}^{\infty} \Knuabs \Bnu  \ d\nu = \int\limits_{0}^{\infty} \Knuabs \Jnu \ d\nu \ . \label{eq:LTE}
\end{equation}
These equations are solved for a spherically symmetric envelope of inner radius $\Rin$ and outer radius $\Rout$, surrounding a star of radius $\Rstar$.

\subsection{Numerical scheme}

\subsubsection{Finite-difference approximation}

\noindent Following a finite-difference procedure, we discretise and sample in a logarithmic way the frequency domain into $n_\nu$ points, denoted by the subscript $k$. Space is discretised into $n_x$ cells and denoted by the subscript $i$. $\Jnu$ is computed at the cell centres and the vector $\Hnu$ on the walls. The differential operator is approximated with a second-order finite differences operator. The equation is solved with respect to a new variable $x=f(r)$ on a regular grid of constant step $\Delta x = \left(x(\Rout) - x(\Rin)\right)/\left(n_x-2\right)$ to allow $r$ to be non-uniformly sampled. We make use of one ghost cell for each grid border to ensure the BCs. We obtain the following system of equations:
\begin{equation}
 A_{k,i+\frac{1}{2}} J_{k,i+1} - A_{k,i}  J_{k,i} +  A_{k,i-\frac{1}{2}} J_{k,i-1} = - b_{k,i} B_k(T_i) \ . \label{eq:Numerical_scheme}
\end{equation}
The non-linear coefficients $A$ are given by
\begin{equation}
\begin{split}
    A_{k,i\pm\frac{1}{2}} = r_{i\pm\frac{1}{2}}^2 D_{k,i\pm\frac{1}{2}} \left .\frac{dx}{dr}\right|_{i\pm\frac{1}{2}} ~,~ A_{k,i} = A_{k,i+\frac{1}{2}} + b_{k,i} + A_{k,i-\frac{1}{2}}  \\
    ,~b_{k,i}= \Delta x^2 \left( \left .\frac{dx}{dr}\right|_{i} \right) ^{-1}  r_{i}^2 \kappa^{\mathrm{abs}}_{k,i} \ . \label{eq:coeff}
    \end{split}
\end{equation}
The non-linear nature of the equation arises from the expression of the coefficients $A$ and RHS in Eq.~(\ref{eq:Numerical_scheme}). They implicitly depend on $\Jnu$, through the diffusion coefficient $\Dnu$ and the radiative equilibrium Eq.~(\ref{eq:LTE}), respectively. The coefficients $A$ require an estimation of $\Jnu$ and its gradient (see Eqs.~(\ref{eq:Dnu}) and (\ref{eq:R})) at the cell walls, given by
\begin{equation}
    J_{k,i+\frac{1}{2}} = \frac{1}{2}\left( J_{k,i+1} +  J_{k,i} \right), \ \left. \nabla \Jnu\right|_{k,i+\frac{1}{2}} = \left.\frac{d x}{dr}\right|_{i+\frac{1}{2}}\frac{J_{k,i+1} -  J_{k,i}}{\Delta x} \ \hat{r} \ . \label{eq:finitediffapprox}
\end{equation}

\subsubsection{Iterative scheme}

Several strategies are possible in order to solve the FLD Eq.~(\ref{eq:Numerical_scheme}) coupled with Eq.~(\ref{eq:LTE}). The simplest approach is to use an iterative method to fully solve Eq. ~(\ref{eq:Numerical_scheme}) and to update the temperature through Eq.~(\ref{eq:LTE}). Iterating between these two processes until convergence yields the solution of the problem. This procedure, commonly called the $\Lambda$-iteration in the literature \citepads[see]{2002ApJ...569..846Y}, becomes very slow and does not converge for large optical depths. In analogy with the usual accelerated $\Lambda$-iteration (ALI) methods, \citetads{2002ApJ...569..846Y} found an improved convergence behaviour by splitting the solution of equation Eq.~(\ref{eq:1D_FLD}) in Eq.~(\ref{eq:LTE}). 

We found a simple method, inspired by the Gauss-Seidel approach, to solve Eqs.~(\ref{eq:Numerical_scheme}) and (\ref{eq:LTE}) simultaneously instead of repetitively. If we denote by $n$ the iteration of the method, we have
\begin{equation}
    J_{k,i}^{n+1} = \frac{b_{k,i} B_k(T_i^n) + A_{k,i+\frac{1}{2}}^n J_{k,i+1}^n + A_{k,i-\frac{1}{2}}^n J_{k,i-1}^{n+1}}{A_{k,i}^n} \ , \label{eq:GS}
\end{equation}
and the temperature is updated after only one Gauss-Seidel spatial sweep at each frequency through Eq.~(\ref{eq:LTE}), which we rewrite as
\begin{equation}
    \sum_{k=0}^{n_\nu -1 } W_k~\kappa^{\mathrm{abs}}_{k,i} B_k(T_i^{n+1}) = \sum_{k=0}^{n_\nu -1 } W_k~\kappa^{\mathrm{abs}}_{k,i} J_{k,i}^{n+1} \ . \label{eq:LTE_quadrature}
\end{equation}
We replaced the frequency integration by a quadrature formula with the associated weights $W_k$. The left-hand side (LHS) of Eq.~(\ref{eq:LTE_quadrature}) was pre-computed and stored in a table, for a wide range of temperatures, allowing the RHS to be linearly interpolated in this table (in the logarithm of the integral for better accuracy) . By doing so, we avoided using a Newton-Raphson procedure to determine the new temperature, which reduces the computational time. 

Our procedure consisted of repeatedly updating $\Jnu$ and $T$ with the help of Eqs.~(\ref{eq:GS}) and (\ref{eq:LTE_quadrature}) until we reached convergence. The coefficients $A$ Eq.~(\ref{eq:coeff}) and the BCs Eq.~(\ref{eq:BC_num}) were immediately updated for each frequency $k$ after one Gauss-Seidel spatial sweep. We note that this procedure is different from the usual $\Lambda$-iteration presented above because the temperature is updated simultaneously with $\Jnu$, within the same iteration $n$.

\subsection{Update of the stellar temperature}

The radiative equilibrium inside the inner cavity requires updating the stellar temperature (see Sect.~\ref{subsect:warming}). Following Eq.~(\ref{eq:teff_update}), we updated the stellar temperature at the end of each iteration $n$ of the numerical scheme,
\begin{equation}
    \Tstar^{n+1} = \left( \Teff^4 - \frac{1}{\sigma} \left(\frac{\Rin}{\Rstar}\right)^2 \sum_{k=0}^{n_\nu -1} W_k \  F_k^{\mathrm{fall},n+1} \right)^{\frac{1}{4}}  \ .
\end{equation}
Here again, we replaced the frequency integration by a quadrature formula with the associated weights $W_k$. $F_k^{\mathrm{fall},n+1}$ is given by Eq.~(\ref{eq:Ffall}) and computed with the freshly updated values of $J_{k,\frac{1}{2}}^{n+1}$ and $R_{k,\frac{1}{2}}^{n+1}$.

\subsection{Boundary conditions}

We used two ghosts cells (one at each boundary of the domain) in order to simplify the implementation of our BCs. In doing so, the inner BC was imposed at the wall between the first ($i=0$) and second cell ($i=1$), and at the outer BC between the last ($i=n_x -1$) and penultimate cell ($i=n_x-2$). As indicated by Eqs.~(\ref{eq:BC_in}) and (\ref{eq:BC_out}), we need to specify $\Jnu$ and $\vec{\nabla} \Jnu$ at these interfaces. For this, we used Eq.~(\ref{eq:finitediffapprox}) and write
\begin{equation}
    \begin{split}
        \Jnu(\Rin|\Rout) \approx \frac{J_{k,0|n_x-2} + J_{k,1|n_x-1}}{2} \ , \\
        \hat{s}.\vec{\nabla} \Jnu |_{\Rin|\Rout} \approx \left.\frac{dx}{dr}\right|_{\frac{1}{2}|n_x -\frac{3}{2}}~ \frac{J_{k,1|n_x-1} - J_{k,0|n_x-2} }{\Delta x} \ .
    \end{split}
\end{equation}
Accordingly, the values of $\Jnu$ in the ghosts cells were updated immediately after one Gauss-Seidel sweep Eq.~(\ref{eq:GS}) to ensure the BCs,
\begin{equation}
    \begin{split}
        J_{k,0} = \frac{ 2 \Delta x \left(\frac{\Rstar}{\Rin}\right)^2 B_k(\Tstar) -  \left[ \Delta x \left(\alpha_{k} - \gamma_{k} \right) - 2 \frac{dx}{dr}|_{\frac{1}{2}} \beta_{k}  \right] J_{k,1} }{\Delta x \left(\alpha_{k} - \gamma_{k} \right) + 2 \frac{dx}{dr}|_{\frac{1}{2}} \beta_{k}} \ , \\
        J_{k,n_x-1} = \frac{ \left[ 2 \frac{dx}{dr}|_{n_x -\frac{3}{2}} \zeta(R_k^\mathrm{e}) D_k(R_k^\mathrm{e}) - \Delta x  \right] J_{k,n_x - 2} }{2 \frac{dx}{dr}|_{n_x -\frac{3}{2}} \zeta(R_k^\mathrm{e}) D_k(R_k^\mathrm{e}) + \Delta x} \ , \label{eq:BC_num}
    \end{split}
\end{equation}
with $\alpha_{k}$, $\beta_{k}$ and $\gamma_k$ being defined by Eqs.~(\ref{eq:alpha}) and (\ref{eq:gamma}). For the extrapolated value $R_k^\mathrm{e}$ in $\zeta(R_k^\mathrm{e})$ Eq.~(\ref{eq:zeta}) and $D_k(R_k^\mathrm{e})$, we used a second-order Lagrange extrapolation, 
\begin{equation}
    R_k^\mathrm{e} = 3 \left( R_{k,n_x -\frac{5}{2}} - R_{k,n_x -\frac{7}{2}}  \right) + R_{k,n_x -\frac{9}{2}} \ . \label{eq:Rnuextrapolation}
\end{equation}

\subsection{Initial conditions}

Initial conditions for both $\Jnu$ and $T$ must be provided in order to solve Eq.~(\ref{eq:Numerical_scheme}). It is clear that the overall convergence speed strongly depends on the initial setup of the solution, but there is also a trade-off with the stability, that is, the ability of the solution to converge, for a wide variety of cases. As an initial guess, we used the analytic solution of the FLD in the optically thin limit Eq.~(\ref{eq:FLD_thin}), and we write
\begin{equation}
    J_{k,i}^0 = \frac{1}{4}\left(\frac{\Rstar}{r_i}\right)^2 B_k(\Tstar) \ ,
\end{equation}
from which we deduce the corresponding temperature profile $T^0$ with the help of Eq.~(\ref{eq:LTE_quadrature}). 
\section{Numerical tests: spherically symmetric envelopes} \label{sect:1D_tests}

\subsection{Benchmarks from Ivezic et al.~(1997)} \label{subsect:FLD_vs_Ivezic}

\begin{figure*}
    \centering
    \includegraphics[width=17cm]{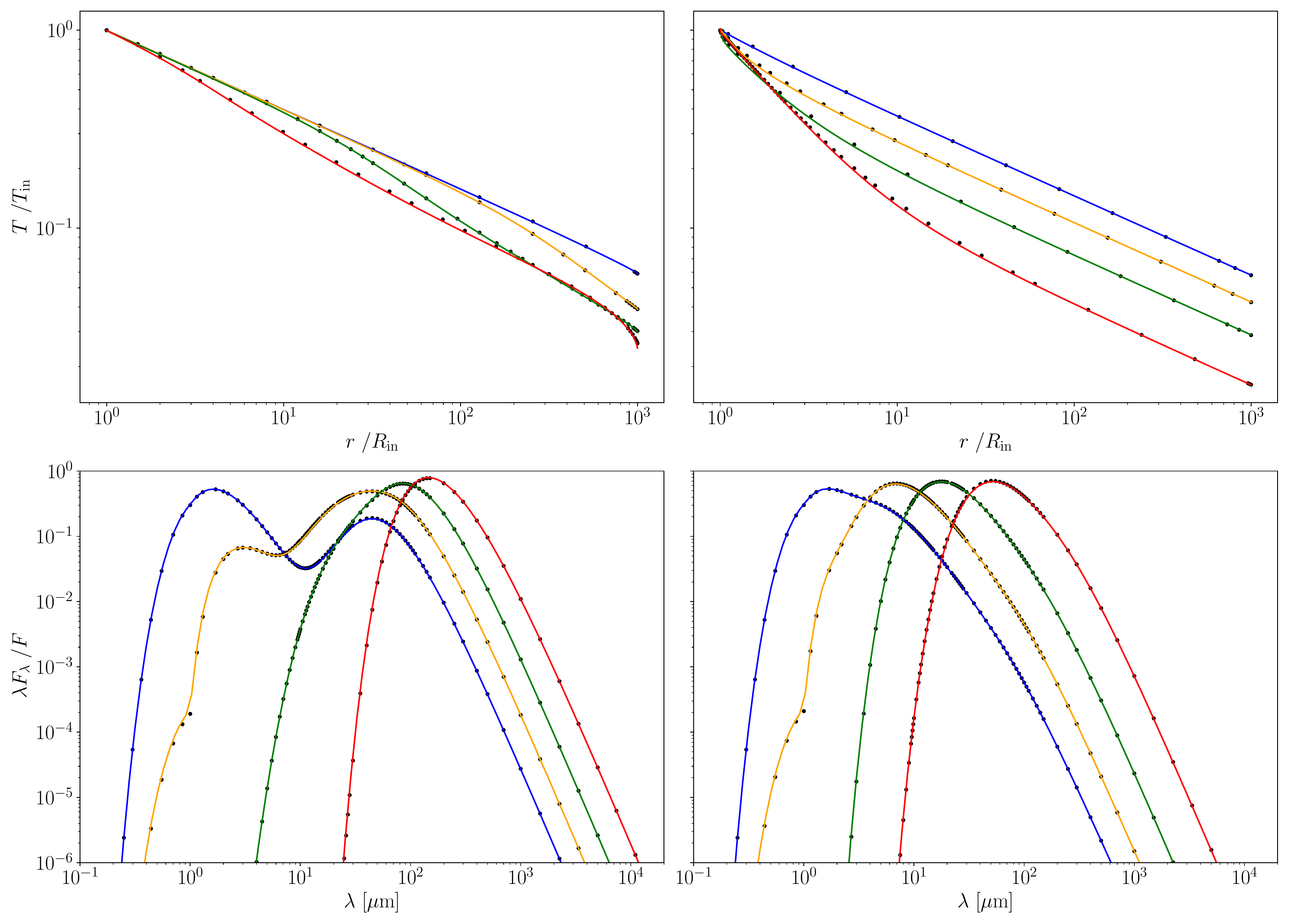}
    \caption{Non-grey case : Normalised temperature profiles (\textit{upper panels}) and SEDs (\textit{lower panels}) for four different opacities $\tau_{\nu_0} = 1, 10, 100, \text{ and } 1 \, 000$ (blue, orange, green and red, respectively) and two density power laws: $p = 0$ (\textit{left panels}) and $p = 2$ (\textit{right panels}). The solid lines represent the FLD curves, and the black dots indicate the benchmark profiles from \citetads{1997MNRAS.291..121I}.}
    \label{fig:FLD_vs_Ivezic}
\end{figure*}

\begin{table}
    \centering
    \caption{Results from the comparison with DUSTY}
    \begin{tabular}{c c c c c}
        \hline\hline
        $\tau_{\nu_0}$ & \multicolumn{2}{c}{$\epsilon(T)$}  & \multicolumn{2}{c}{$\epsilon\left(\lambda F_\mathrm{\lambda} / F\right)$} \\
         & $p = 0$ & $p = 2$ & $p = 0$ & $p = 2$ \\
        \hline
        1 & 0/0/1 & 0/0/1 & 1/1/4 & 1/1/3  \\
        10 & 0/0/1 & 1/1/3 & 2/3/14 & 2/2/10  \\
        100 & 1/0/1 & 1/1/3 & 1/1/8 & 1/2/11  \\
        1~000 & 2/1/4 & 1/1/4 & 6/8/30 & 3/2/9  \\
        \hline
    \end{tabular}
    
    \tablefoot{Relative differences $\epsilon$ (in \%) for the temperature profiles $\epsilon(T)$ and for the SEDs $\epsilon\left(\lambda F_\mathrm{\lambda} / F\right)$ shown in Fig.~\ref{fig:FLD_vs_Ivezic}. The results are presented in the form mean($\epsilon$) / std($\epsilon$) / max($\epsilon$) and rounded to the closest percent.}
    
    \label{tab:FLD_vs_Ivezic}
\end{table}

We tested the accuracy of our FLD code with our Robin-type mixed boundary conditions in a general and realistic case, by comparing it with the 1D benchmark problems realised by \citetads{1997MNRAS.291..121I}. We recall the conditions of the test and refer to the original paper for further information. 

A point source surrounded by a spherically symmetric envelope of matter at radiative equilibrium irradiates as a black body at the temperature $\Tstar = 2 \, 500 \ \mathrm{K}$. This envelope extends from the inner radius $\Rin$ to the outer radius $\Rout = 1 \, 000 \ \Rin$. The inner radius is set so that the temperature at the inner radius is always $\Tin = T(\Rin)= 800 \ \mathrm{K}$. The density profile $n(r)$ is assumed to be a power law of the form $n(r) = n_0 \left(\Rin/r\right)^{p}$. The radial optical depth $\tau_\nu$ of the envelope is linked to the density profile by
\begin{equation}
    \tau_{\nu} = \int\limits_{\Rin}^{\Rout} \Knuext \ dr = \int\limits_{\Rin}^{\Rout} \Cnuext \ n(r) \  dr \ ,
\end{equation}
where $\Cnuext$ is the extinction cross-section coefficient. It is defined by
\begin{equation}
    \begin{array}{ll}
        \Cnuabs = C_{\nu_0}^{\mathrm{abs}} \ , \  \Cnusca = C_{\nu_0}^{\mathrm{sca}} & \text{if} ~~ \nu \geq \nu_0 \ ,  \\
        \Cnuabs = C_{\nu_0}^{\mathrm{abs}} \left(\frac{\nu}{\nu_0}\right) \ , \ \Cnusca = C_{\nu_0}^{\mathrm{sca}} \left(\frac{\nu}{\nu_0}\right)^4 & \text{if} ~~ \nu \leq \nu_0 \ , \\
        \Cnuext = \Cnusca + \Cnuabs \ ,
    \end{array}
\end{equation}
with $C_{\nu_0}^{\mathrm{abs}} = (1-\eta) \, C_{\nu_0}^{\mathrm{ext}}$, $C_{\nu_0}^{\mathrm{sca}} = \eta \, C_{\nu_0}^{\mathrm{ext}}$, $\nu_0$ the frequency corresponding to $\lambda = 1 \ \mathrm{\mu m}$ and $\eta$ the albedo, set to $1/2$ for these tests. The benchmark problems are thus completely defined by two parameters: (i) the exponent in the density power law  $p = 0, 2$, and (ii) the radial optical depth of the envelope at $\nu_0$, $\tau_{\nu_0} = 1, 10, 100, \text{ and } 1 \, 000$. This created eight different cases to test the accuracy of our code. The coefficient $n_0$ in the density profile is derived with the help of $\tau_{\nu_0}$ and $p$,
\begin{equation}
    n_0 =  \frac{\left(p-1\right) \tau_{\nu_0}}{C^{\mathrm{ext}}_{\nu_0} \Rout} \left( \frac{\Rout}{\Rin}\right)^p \left[ \left(\frac{\Rout}{\Rin}\right)^{p-1} -1 \right]^{-1}.
\end{equation}
The normalised temperature profile $T / \Tin$ and the normalised SED $\lambda F_\lambda / F$ ($F = \int_0^\infty F_\lambda \ d\lambda$) of the envelope are shown in Fig.~\ref{fig:FLD_vs_Ivezic} for each case. 

The Ivezic benchmarks were produced with version 2 of DUSTY\footnote{available at \href{http://faculty.washington.edu/ivezic/dusty_web/}{http://faculty.washington.edu/ivezic/dusty\_web/}} (\citeads{1997MNRAS.287..799I}). Because our code is of different nature, the spatial and frequency grids are different. We then compared the results by linearly interpolating our profiles (in $\log-\log$ scale) on the DUSTY grids. We used 128 points for space and frequency, with a logarithmic sampling. The corresponding relative differences are displayed in Table \ref{tab:FLD_vs_Ivezic}. We also point out that we restricted the comparison, for the normalised SEDs, to the frequency domain where $\lambda F_\mathrm{\lambda} / F \geq 10^{-6}$ because of non-physical results of the DUSTY code below this threshold, for the smallest wavelengths.

The FLD results and the benchmarks agree well. The average relative differences in the temperature profiles $\mean(\epsilon(T))$ is of the order of $1 \ \%$, with a maximum value of approximately $4 \ \%$, achieved by the most optically thick envelopes ($\tau_{\nu_0} = 1\,000$). The average of the relatives differences in the normalised SEDs $\mean(\epsilon\left(\lambda F_\mathrm{\lambda} / F\right))$ always stays below $3 \ \%$, with the exception of the optically thick envelope with constant density profile, where this difference reaches $6 \ \%$.

\subsection{Grey spherical shell with the Monte Carlo code from Niccolini \& Alcolea (2006)} \label{subsect:FLD_vs_TAPAS}

\begin{figure*}
    \centering
    \includegraphics[width=17cm]{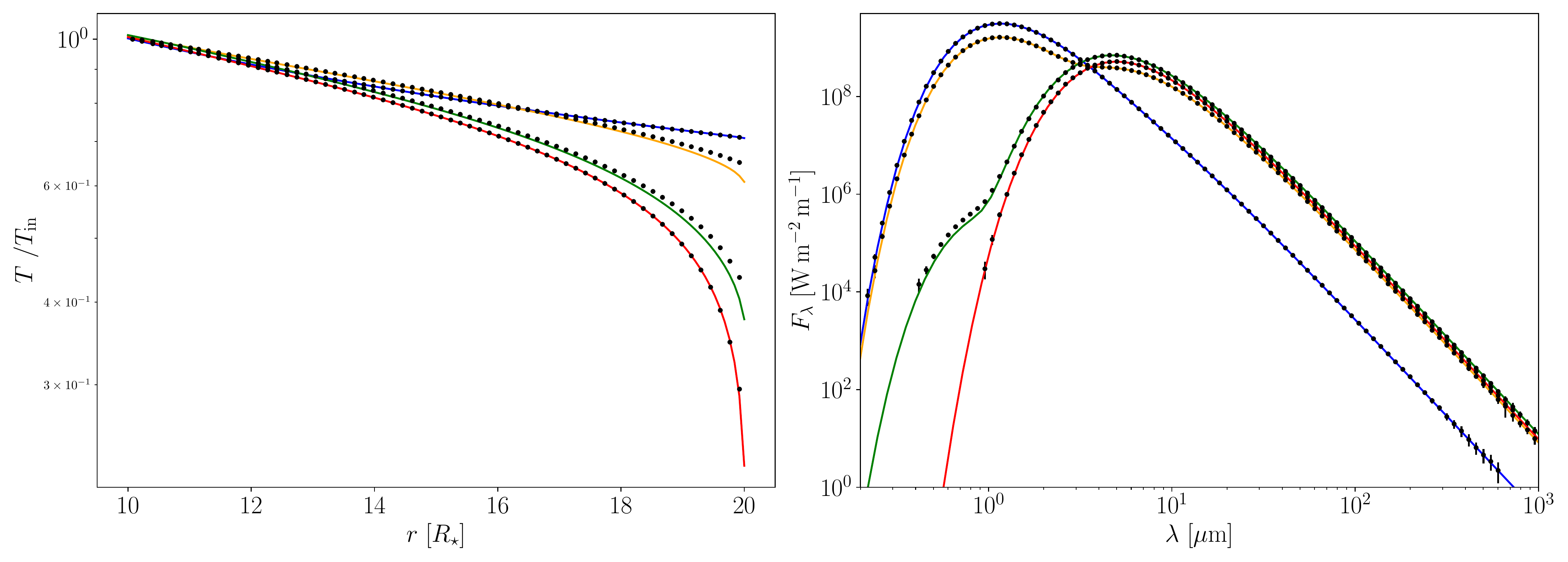}
    \caption{Grey case : Normalised temperature profiles (\textit{left panel}) and emerging fluxes (\textit{right panel}) for four different opacities $\tau = 0.01, 1, 10, \text{ and } 100$ (blue, orange, green and red, respectively) with a constant density profile (p = 0). The solid lines represent the FLD curves, and the black dots with the error bars $\sigma$ indicate the MC profiles from \citetads{2006A&A...456....1N}.}
    \label{fig:FLD_vs_TAPAS}
\end{figure*}

\begin{figure*}
    \centering
    \includegraphics[width=17cm]{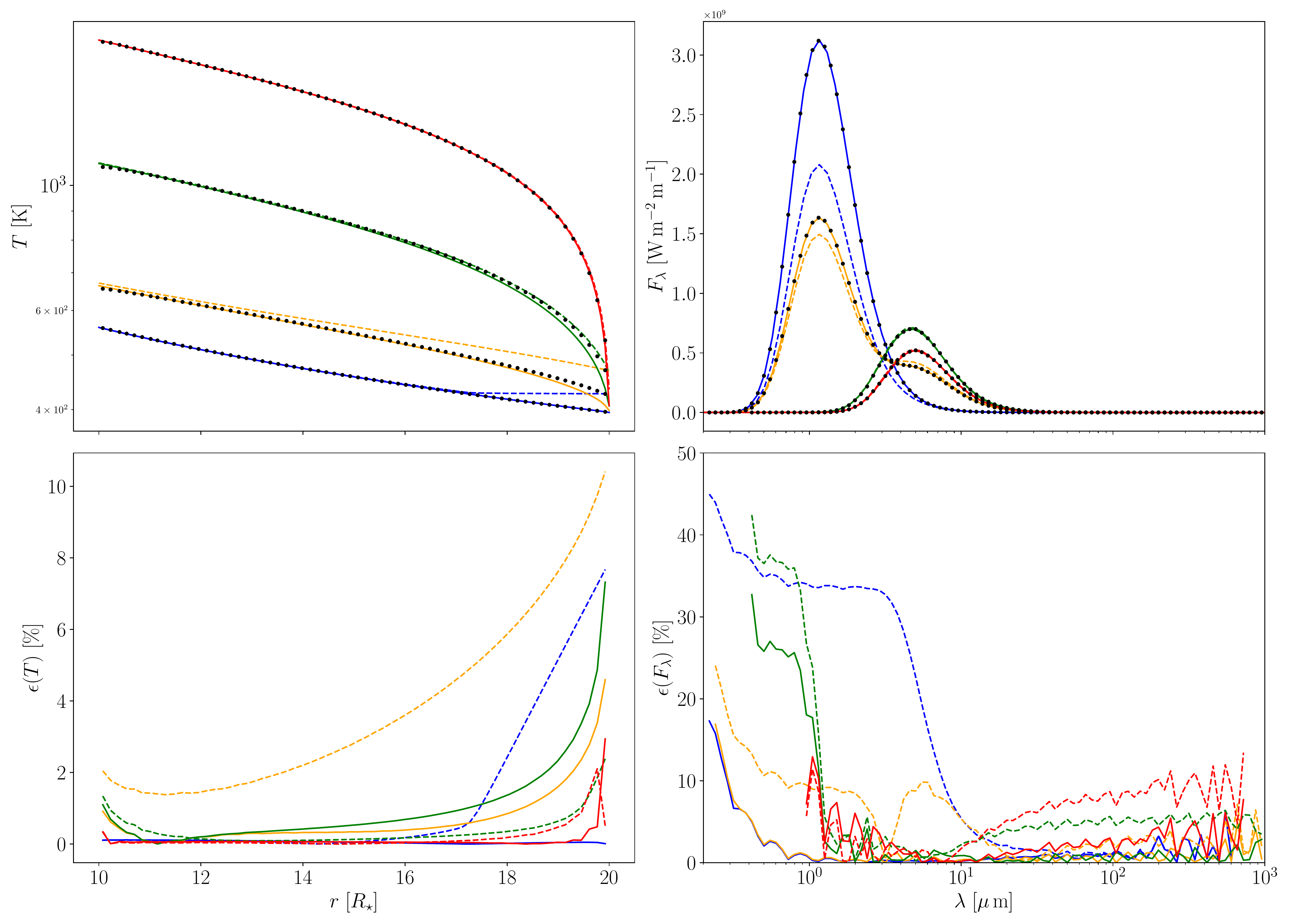}
    \caption{Comparison of the outer boundary conditions presented here (solid lines) and from \citetads{1981ApJ...248..321L} (dashed lines) for the test case presented in Sect.~\ref{subsect:FLD_vs_TAPAS}. The colour code is the same as in Fig.\ref{fig:FLD_vs_TAPAS}. The emerging fluxes are displayed in semi-log scale to highlight the differences between the BCs. The lower panels show the relative differences profiles with respect to the MC code from \citetads{2006A&A...456....1N} in temperature (\textit{left}) and in emerging flux (\textit{right}).}
    \label{fig:comp_bc}
\end{figure*}

\begin{table}
    \centering
    \caption{Results from the comparison with the MC code}
    \begin{tabular}{c c c}
        \hline\hline
        $\tau$ & $\epsilon(T)$ & $\epsilon\left( F_\mathrm{\lambda}\right)$ \\
        \hline
        0.01 & 0/0/0 & 0/3/17 \\
        1 & 1/1/5 & 0/3/17 \\
        10 & 1/1/7 & 1/8/33 \\
        100 & 0/0/3 & 1/2/13  \\
        \hline
    \end{tabular}
    \tablefoot{Relative differences $\epsilon$ (in \%) for the temperature profile $\epsilon(T)$ and for emerging flux $\epsilon\left( F_\mathrm{\lambda} \right)$ shown in Fig. \ref{fig:FLD_vs_TAPAS}. The differences are presented in the form mean($\epsilon$) / std($\epsilon$) / max($\epsilon$) and rounded to the closest percent. mean($\epsilon$) is given by Eq.~(\ref{eq:mean_esp}).}
    \label{tab:FLD_vs_TAPAS}
\end{table}

We compared the FLD code with a 3D MC radiative transfer code \citepads{2006A&A...456....1N}. We wished to test the new BCs in the less realistic but more extreme case of a spatially small spherically symmetric grey envelope. We expect the boundary effects to play a major role for this type of problems. The inner radius was set to $\Rin = 10 \ \Rstar$ and the outer radius $\Rout = 20 \ \Rstar$. We assumed a constant density profile ($p = 0$) in the envelope. Our test cases consisted of determining the normalised temperature profiles $T / T_{\mathrm{in}}$ and the emerging fluxes $F_\lambda$ for several cases, ranging from optically thin ($\tau=0.01$) up to the optically thick envelopes ($\tau=100$). The corresponding profiles are shown in Fig.~\ref{fig:FLD_vs_TAPAS}.

Because the codes are different, we interpolated our results linearly (in $\log-\log$ scale) on the MC grids. The relative differences between the two codes, are displayed in Table \ref{tab:FLD_vs_TAPAS}. As an additional feature, the MC code also provides an estimation of the errors on the temperature $\sigma(T)$ and on the emerging flux $\sigma(F_\lambda)$, computed from the MC noise \citepads{2006A&A...456....1N}. We used this information to compute a more relevant mean value for $\epsilon(T)$ and $\epsilon(F_\lambda)$,
\begin{equation}
    \mean(\epsilon(T)) = \frac{\sum\limits_{i=0}^{N_x} W_i \ \epsilon(T_i) }{\sum\limits_{i=0}^{N_x} W_i} \ , \ \mean(\epsilon(F_\lambda)) = \frac{\sum\limits_{k=0}^{N_\lambda} W_k \ \epsilon(F_k) }{\sum\limits_{k=0}^{N_\lambda} W_k} \label{eq:mean_esp}
\end{equation}
where $N_x$ ($N_\lambda$) is the number of spatial (wavelength) points of the MC grid, $\epsilon(T_i)$ ($\epsilon(F_k)$) is the relative error (in $\%$) on the temperature (emerging flux) between our results and the MC results, and $W_i$ ($W_k$) is the inverse square of the MC relative errors, defined as
\begin{equation}
    W_i = \left( \frac{\sigma(T_i)}{T_i} \right)^{-2} \ , \  W_k = \left( \frac{\sigma(F_k)}{F_k} \right)^{-2}
\end{equation}
The two results agree well. The average of the relative differences of the temperature profile $\mean(\epsilon(T))$, remains of the order of $1 \ \%$ for all the cases we tested. The largest differences are reached for the intermediate cases ($\tau = 1, 10$) and are located on the external edge of the envelope. This is expected because the FLD approximation is known to perform well in the optically thin and thick regimes, but it is less well suited to describe these intermediate cases. Nevertheless, the temperature profile is still quite well reproduced and the emerging flux is not affected by the small errors on the temperature close to the outer edge. We point out that the BCs derived in this paper allow us to successfully reproduce the correct behaviour of the temperature profile for the optically thick envelope where it shows a quite steep decrease at the outer surface, as shown by Fig.~\ref{fig:FLD_vs_TAPAS}. The emerging fluxes $F_\lambda$ agree within $1 \ \%$ on average, except for the optically thick case ($\tau = 100$), where $\mean(\epsilon(F_\lambda))$ reaches about $5 \ \%$. 

We conclude this section with a comparison of the outer BC from this study Eq.~(\ref{eq:BC_out}) and the original BC Eq.~(56) from L\&P, given by, 
\begin{equation}
\Jnu(\Rout) + 2 \Dnu \shat . \vec{\nabla} \Jnu|_{r = \Rout} = 0 \label{eq:L&Pbc}
\end{equation}
We previously mentioned in Sect.~\ref{subsect:inner_cavity} that the inner BC Eq.~(\ref{eq:BC_in}) is analytically identical to that of L\&P in spherical symmetry, so we restrict the comparison to the outer edge of the envelope. It is important to notice that the BC of $L\&P$ was not originally intended to describe this class of problems, but the comparison still remains instructive for studying the importance of the BCs for the accuracy of the solution. In Eq.~(\ref{eq:L&Pbc}), the factor $2$ (Eq.~58 in L\&P) was originally used to give the correct ratio of the energy density over the emerging radiative flux in plane-parallel geometry for an optically thin slab illuminated by an isotropic incident radiation field. However, for the special case of a spherical envelope surrounding a black-body star, this ratio becomes
\begin{equation}
    \frac{\int_{\mu_0}^1 B_\nu(\Tstar) d\mu }{\int_{\mu_0}^1 \mu B_\nu(\Tstar) d\mu} =  \frac{2\left(1-\mu_0\right)}{1-\mu_0^2} = \frac{2}{1 + \mu_0}
\end{equation}
with $\mu_0 = \sqrt{1 - \left(\frac{\Rstar}{\Rout}\right)^2}$, the cosine of the stellar angular size at $\Rout$. As $\mu_0 \rightarrow 0$ or equivalently $\Rstar / \Rout \rightarrow 1$, this ratio increases to $2$, as expressed by L\&P, because we recover the case of a plane-parallel geometry with an isotropic incident radiation field. Far from the star ($\mu_0 \rightarrow 1$ or $\Rstar / \Rout \rightarrow 0$), the ratio tends to $1$, associated with an incoming sharp-peaked radiation. Hence, the L\&P BC that set the ratio to $2$ will strongly deviate from the analytic limit $1$, in the optically thin regime.
In Fig.~\ref{fig:comp_bc} we display the relative differences in the temperature profiles $\epsilon(T)$ and in the emerging fluxes $\epsilon(F_\lambda)$ for the same test case as presented at the beginning of this section. We note that this test case is not realistic, however,  it allows to compare different optical regimes, in contrast to a more realistic problem in which the radiation is free in the external regions most of the time, such as for the test case presented in Sect.~\ref{subsect:FLD_vs_Ivezic}. We recall that the BC from L\&P is a limiting case of the BC derived in this study, in the case $R\ll 1$, where the emerging radiation is almost isotropic (see Sect.~\ref{subsect:outerBC}), hence it is not surprising that the results converge to the same profile, for a optically thick grey envelope ($\tau = 100$). On the other-hand, in the optically thin case ($\tau = 0.01$) where we would expect the $\zeta$ coefficient Eq.~(\ref{eq:zeta}) to be close to unity, the BC from L\&P performs poorly as expected. For intermediates regimes ($\tau = 1, 10$), the entanglement of the error of the BC and FLD method itself makes any comparison very hard. We note that although the temperature profile is closer to the benchmark result for $\tau = 10$ with the BC of L\&P, this is not the case for the associated emerging flux. We also note that the BC, although defined locally, can have a global effect on the whole solution, as shown by the temperature profile of the test case $\tau = 1$. To conclude, we also point out that we tried to implement the L\&P BC in the non-grey cases (Sect.~\ref{subsect:FLD_vs_Ivezic}) but we were unable to reach a satisfying convergence of the computations. 

\section{Conclusion} \label{sect:conclusion}

The FLD approximation together with the new BCs, yields promising results in correctly describing the radiation transport inside spherically symmetric circumstellar envelopes. These conditions, derived in Sect.~\ref{sect:Boundary_Conditions} from physically consistent constraints on the behaviour of the radiation field at the inner and outer surfaces, allow us to compute with a good accuracy the temperature profile and the SED for a wide range of configurations, from very small to very large optical thicknesses. As shown in Sect.~\ref{sect:1D_tests}, it reproduces the correct temperature profile within $\leq$ 2\%, and the SED or emerging flux at less than $\leq 6\%$ on average, with respect to the solution of the full radiative transfer equation under radiative equilibrium. 

The numerical solution of the 1D non-linear diffusion equation Eq.~(\ref{eq:1D_FLD}) coupled with the radiative equilibrium Eq.~(\ref{eq:LTE}) was performed with a Gauss-Seidel method-based iterative scheme, in which the temperature is updated at each iteration step. Furthermore, we point out that the FLD approximation implemented with the proposed outer-boundary condition provides a simple approximation for the flux emitted by the envelope. This allows computing the SED or emerging flux without using any ray-tracing module, which is a significant gain in computational time.

The next step, which will be our main concern for our future work, is the generalisation of these boundary conditions to non-spherically symmetric media, in particular for circumstellar discs, where the BCs will take a more complex form. In this regard, the extension to 2D will require numerical optimisations. Several directions of improvements are already being studied, such as the use of multi-frequency adaptive spatial grids, or acceleration procedures for efficiently solving the FLD equation under the radiative equilibrium condition. The last point is also crucial for solving extreme optically thick envelopes.

\begin{acknowledgements}
This study has been supported by the Lagrange laboratory of Astrophysics and funded under a 3-year PhD grant from Ecole Doctorale Sciences Fondamentales et Appliquées (EDSFA) of the Université Côte-d'Azur (UCA). The benchmark profiles were generated with the radiative transfer code DUSTY, available at \hyperlink{http://faculty.washington.edu/ivezic/dusty_web}{http://faculty.washington.edu/ivezic/dusty\_web}. Part of the computations were carried out with the help of OPAL-Meso computing facilities. The authors are grateful to the OPAL infrastructure from
Observatoire de la Côte d'Azur (CRIMSON) for providing resources
and support.
\end{acknowledgements}


\bibliographystyle{aa} 
\bibliography{references} 

\end{document}